\documentclass[twocolumn]{aastex62}
\usepackage{amsmath,amstext}
\usepackage[T1]{fontenc}
\usepackage{apjfonts} 
\usepackage[figure,figure*]{hypcap}

\newcommand{\enzo}{\texttt{ENZO}}
\newcommand{\moray}{\texttt{MORAY}}
\newcommand{\Ms}{{\ensuremath{M_{\odot} }}}
\newcommand{\Zs}{\ensuremath{Z_\odot}}

\shorttitle{Pop III IMF and the First Galaxies}
\shortauthors{Chen et al.}

\begin{document}

\title{How the Population III Initial Mass Function Governs the Properties of the First Galaxies}

\author{Li-Hsin Chen}
\affiliation{Institut f\"{u}r Theoretische Astrophysik, Zentrum f\"{u}r Astronomie, Universit\"{a}t Heidelberg, Heidelberg 69120, Germany}
\affiliation{Institute of Astrophysics, National Taiwan University,  Taipei 10617, Taiwan} 
\affiliation{Institute of Astronomy and Astrophysics, Academia Sinica,  Taipei 10617, Taiwan} 

\author{Ke-Jung Chen}
\affiliation{Institute of Astronomy and Astrophysics, Academia Sinica,  Taipei 10617, Taiwan} 

\author{Sung-Han Tsai}
\affiliation{Institute of Astrophysics, National Taiwan University,  Taipei 10617, Taiwan}
\affiliation{Institute of Astronomy and Astrophysics, Academia Sinica,  Taipei 10617, Taiwan} 
\affiliation{Institute of Astronomy, National Central University, Taoyuan 32001, Taiwan}

\author{Daniel Whalen}
\affiliation{Institute of Cosmology and Gravitation, University of Portsmouth, Portsmouth PO1 
3FX, UK} 
\affiliation{Ida Pfeifer Professor, University of Vienna, Department of Astrophysics, Tuerkenschanzstrasse 17, 1180, Vienna, Austria}

\correspondingauthor{Li-Hsin Chen}
\email{Li-Hsin.Chen@uni-heidelberg.de}

\begin{abstract}

The properties of Population III (Pop III) stars impact many aspects of primeval structure formation such as the onset of cosmological reionization and early chemical enrichment.  However, in spite of over twenty years of numerical simulations and attempts to constrain the Pop III initial mass function (IMF) by stellar archaelogy, little is known of the masses of the first stars for certain.  Here, we model the effect of Pop III IMF on the properties of primeval galaxies with a suite of high-resolution radiation-hydrodynamical simulations with \enzo.  We find that a top-heavy Pop III IMF results in earlier star formation but dimmer galaxies than a more conventional Salpeter-type IMF because explosions of massive Pop III stars produce more turbulence that suppresses high-mass second-generation star formation.  Our models suggest that the Pop III IMF could therefore be inferred from detections of primordial galaxies, which will be principal targets of the {\em James Webb Space Telescope} and extremely large telecopes on the ground in the coming decade.

\end{abstract}


\section{Introduction}
\label{Introduction}

The birth of primordial (Pop III) stars at $z \sim$ 20 - 25 marked the end of the cosmic dark ages and the onset of primeval galaxy formation. Pop III stars are thought to form in primordial halos that have grown to 10$^5$ - 10$^6$ \Ms\ by mergers and accretion, when enough H$_2$ can form to cool gas and create stars. The original numerical simulations of Pop III star formation (SF) suggested that they have typical masses of 100 - 200 \Ms\ and form in isolation, one per halo \citep{bcl01,nu01,abn02}, but later studies have since shown that they can form in binaries \citep{turk09,sb13} or small multiples \citep{stacy10,clark11,get11}.  The final masses of Pop III stars depends on how many of them form in a halo, when their ionizing UV flux halts accretion \citep{tm08,hos11,susa13,hir13,hir15}, and whether or not they are ejected from the disk by gravitational torques \citep{get12}. They have been found to range from less than a solar mass to up to several hundred solar masses.  However, the Pop III IMF remains unknown because primordial stars cannot be directly observed and simulations still lack crucial physics such as magnetic fields and radiation transport \citep[or be run for times that are long enough to reach the main sequence at the required numerical resolution;][]{fsg09,glov12,dw12,greif14}.

Pop III stars radically transform their environments, first by photoevaporating the halos in which they are born \citep{wan04,ket04,awb07} and at least partially ionizing others nearby \citep{su06,hus09}.  Then, depending on their masses, some explode as supernovae (SNe), expelling large masses of the first heavy elements in the universe.  Pop III stars with masses of 8 - 30 \Ms\ die as core-collapse (CC) SNe and 90 - 260 \Ms\ stars explode as highly energetic pair-instability (PI) SNe \citep{hw02,hw10}.  A few very-rapidly rotating 30 - 60 \Ms\ Pop III stars may die as gamma-ray bursts \citep[GRBs; e.g.,][]{mes13a} or hypernovae \citep[HNe; e.g.,][]{smidt13a}.   A number of studies have examined how metals from Pop III SNe propagate into the universe on a variety of scales, from inside the halo itself \citep{slud16} to out into the relic H II region of the progenitor star \citep{get07,ritt12,rit16,chen17b,mag20,taru20,latif20c}.  Both radiation \citep{yoh07,suh09} and metals \citep{mbh03,schn06,ss07,brit15} from the first stars cause subsequent generations of stars to form on smaller mass scales.  Pop III SNe could constrain the masses of the first stars, either directly through their detection in the near infrared \citep[NIR;][]{tet12,tet13,ds13,wet12a,wet12c,wet13c,wet12b,ds14,hart18a,moriya19,ryd20a} or indirectly from their nucleosynthetic imprint on less massive second-generation stars that may still live today \citep[e.g.,][]{fet05,bc05,iet05,jet09b,jw11,frebel10,chen17a,chen17b,ish18,hart18b,hart19a}.

When primordial halos grow to 10$^8$ - 10$^9$ \Ms\ by mergers and accretion they are massive enough to maintain consecutive cycles of stellar birth and explosion without all the fuel for forming new stars being blown into the IGM, becoming, in effect, primitive galaxies \citep{get08,fg11}.  Ionizing UV, winds and SNe from stars in these galaxies regulate the rise of later generations and determine their observational signatures \citep{get10,jeon11,pmb12,wise12,ss13,jeon14,corl18,jeon19a}.  The simulations of primordial galaxies performed to date follow their formation in cosmological environments from early times but assume a fixed IMF for Pop III and Pop II stars that does not evolve over time \citep[e.g.,][]{ren15,fiby15}.  These campaigns produce galaxies with realistic features but cannot parametrize their observables as a function of IMF.  

Detections of primordial galaxies by the {\em James Webb Space Telescope} \citep[{\em JWST};][]{jwst,jwst2} and 30 - 40 m telescopes on the ground in the coming decade could constrain the properties of the first stars if the spectra of primordial galaxies can be linked to a specific Pop III IMF.  Here, we investigate how the properties of the first galaxies vary with the number and type of Pop III SNe. In Section 2 we describe our numerical methods and protogalaxy models.  We present the results of our simulations in Section 3 and conclude in Section 4.

\section{Numerical Method}
\label{Methods}

We model the formation of primordial galaxies with the \enzo\ adaptive mesh refinement (AMR) cosmology code \citep[v2.5;][]{enzo}.  \enzo\ utilizes an adaptive particle-mesh $N-$body scheme \citep{efs85,couch91} to evolve dark matter (DM) and a third-order accurate piecewise-parabolic method for fluid flows \citep{wc84,bryan95}.  We use the low-viscosity Harten-Lax-van Leer-Contact (HLLC) Riemann method \citep{toro94} for capturing strong shocks and rarefaction waves to prevent negative energies or densities in the simulation.  \enzo\ self-consistently evolves nine-species non-equilibrium gas chemistry with hydrodynamics \citep[H, H$^+$, e$^-$, He, He$^+$, He$^{++}$, H$^-$, H$_2$ and H$_2^+$;][]{abet97,anet97} and includes primordial gas cooling in the energy equation:  collisional excitational and ionizational cooling by H and He, recombinational cooling, bremsstrahlung cooling and H$_2$ cooling.  We use H$_2$ cooling rates from \citet{ga08} and metal cooling rates from \citet{japp07}.  

\subsection{Protogalaxy Model}

\begin{figure}
\plotone{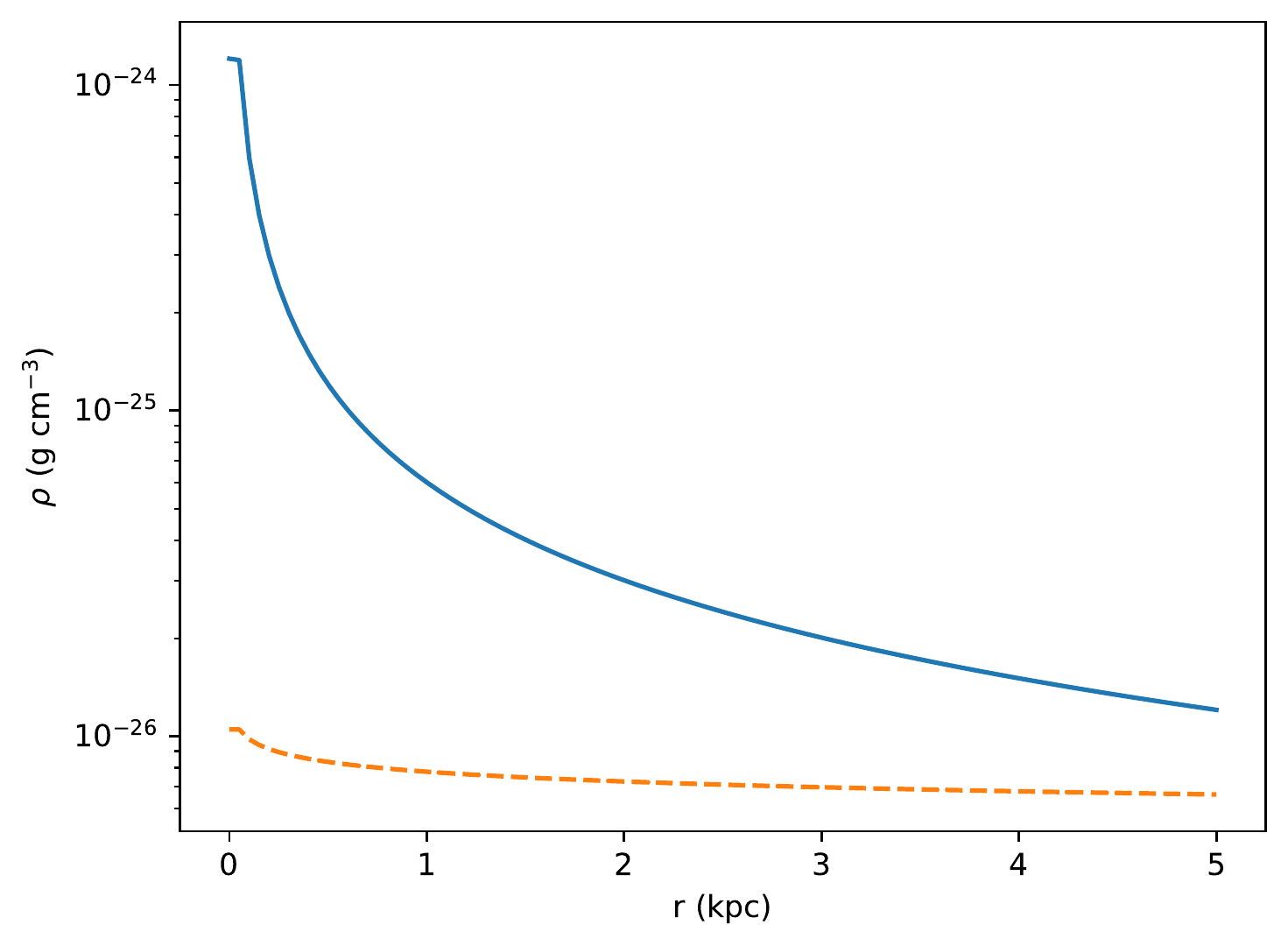}
\caption{Density profiles of the 3 $\times$ 10$^{8}$ \Ms\ halo (gold dashed) and 1 $\times$ 10$^{9}$ \Ms\ halo (blue solid).}
\label{fig:nfw}
\end{figure}

\begin{figure*}
\centering
\includegraphics[width=\textwidth]{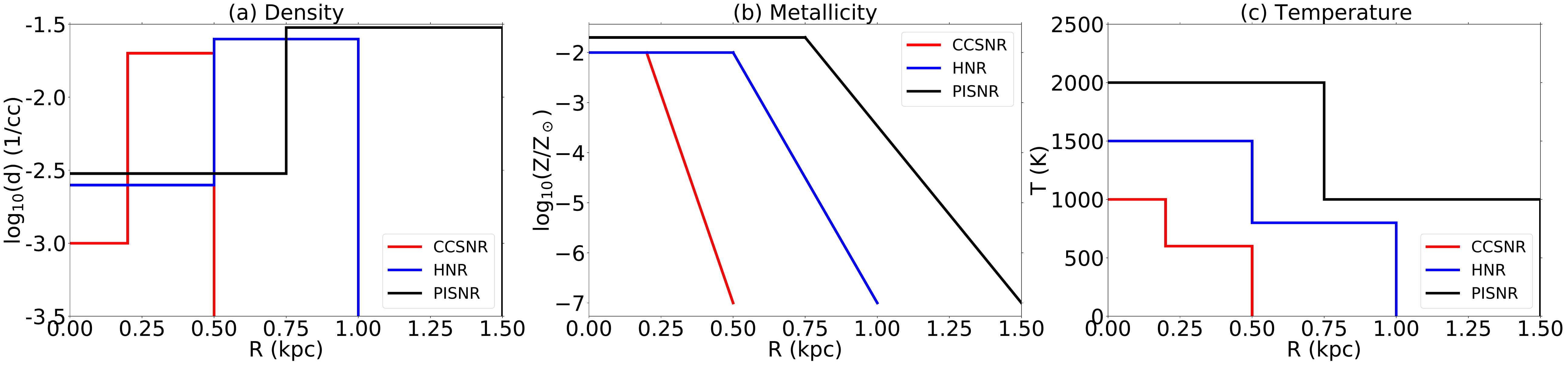}
\caption{Left to right: radial profiles of number density, metallicity, and temperature of the three types of SNRs in our simulations, respectively.}
\label{fig:snrprofile}
\end{figure*}

The protogalaxies in our simulations are approximated as isolated Navarro-Frenk-White (NFW) halos \citep{nfw96}. An NFW profile is used to initialize the gravitational potential due to the DM in the halo but DM itself is not included in our runs.  This potential is static and does not evolve over the run.  We adopt idealized profiles for the initial structure of the protogalaxy instead of evolving them from primordial density fluctuations at high redshifts so we can efficiently run large grids of models and better identify the effect of Pop III IMF on their structures at later times.  We consider two halo masses, 3 $\times$ 10$^8$ \Ms\ and $10^9$ \Ms, which span the minimum masses of the first galaxies. Their corresponding baryon masses, $M_\mathrm{bar}$, are $\sim$ 6 $\times$ 10$^7$ \Ms\ and $\sim$ 1 $\times$ 10$^8$ \Ms, respectively.  We truncate the density profiles of both halos at a radius of 4 kpc, reducing them to the IGM mean value of 1.673 $\times$ 10$^{-27}$ g cm$^{-3}$ (see Figure~\ref{fig:nfw}).

At the beginning of each run, Pop III SN remnants (SNRs) are initialized in the halo as spheres with density, temperature and metallicity profiles taken from explosions in cosmological environments in \citet{chen14c} and \citet{chen15} (see Figure~\ref{fig:snrprofile}).  These profiles do not include the relic H II region of the star, which can extend out to twice the radius of the SNR after it comes into pressure equilibrium in the warm, partially recombined gas and stalls \citep{get07}.  We consider three types of SNe: a 15 \Ms\ CC SN (1.2 foe), a 30 \Ms\ HN (10 foe) and a 200 \Ms\ PI SN (10 foe), where 1 foe $=$ 10$^{51}$ erg.  The metal yields of these explosions are 1.8 \Ms, 26.4 \Ms, and 202.8 \Ms, respectively.  The SNRs are randomly distributed throughout the halo but are all assigned free-fall velocities towards its center.  Because we do not evolve halos in cosmological environments we neglect inverse Compton (IC) cooling in the SNe.  IC cooling losses in Pop III PI SNRs can be up to half of the original energy of the explosion in isolated, low-mass halos \citep{ky05,wet08a} and can therefore limit how far metals from these explosions propagate into the intergalactic medium (IGM).  However, IC losses from Pop III CC SNe are small, and the PI SNe in our halos occur in much deeper potential wells so their remnants are confined to smaller radii that enclose smaller volumes of cosmic microwave background (CMB) photons.  We therefore do not expect IC cooling to strongly affect the propagation of metals in our halos.  For consistency, we also ignore temperature floors in the gas imposed by the CMB ($T$ $=$ 2.73 $(1+z)$ K), but in practice gas never cools to these temperatures in our simulations.
 
The initial number of SNRs in each halo and relative numbers of explosion type are determined by our choice of Pop III IMF.  Here, we consider a Salpeter-like IMF \citep[SAL;][]{sal55} with a peak at 10 \Ms\ and a top-heavy IMF from \citet{hir15} (HIR15).  Not every constituent halo of the protogalaxy hosts a SN so in the 3 $\times$ 10$^8$ \Ms\ halo we assume that there are 5 CC SNRs if we use the SAL IMF and 3 PI SNRs if we use the HIR15 IMF.  In the 10$^9$ \Ms\ halo we initialize 18 CC SNe and 2 HNe if we use the SAL IMF and 2 CC SNe and 7 PI SNe if we use HIR15.  We sample SNR merger scenarios by distributing the remnants randomly throughout the halo or randomly with the exception of one, which is placed at the center and is either a HN or PI SN, depending on IMF.  

Baryon densities in our halos do not precisely trace the NFW DM profiles because the gas is assumed to have already settled in the gravitational potential to some degree by the time we begin to evolve them.  We therefore use an $r^{-0.1}$ baryon density profile in the less massive halo and a steeper $r^{-1.0}$ profile in the more massive halo, consistent with greater settling in its deeper potential well.  In halos with SNRs at their centers, we apply uniform density profiles to approximate gas displaced from the center by the explosion.  However, in all these cases the density profiles are normalized to produce the total baryon mass expected for the given halo mass.  Because the gas does not exactly trace the DM profile, it is not initially in  hydrostatic equilibrium.

Each halo is centered in a 10 $h^{-1}$ kpc box with a 128$^3$ root grid.  This grid by itself cannot resolve the SNRs so prior to the launch of the run we recursively refine the internal structure of the halo with up to ten levels of AMR.  This procedure yields a maximum spatial resolution of 0.077 pc, which is sufficient to resolve the SNRs and the internal structures of molecular clouds in the halo at later times.  The grid is flagged for refinement where the baryon overdensity is 8$\rho_0 N^l$, where $\rho_0 =$ 1.673 $\times$ 10$^{-27}$ g cm$^{-3}$ is the ambient density of the halo, $N =$ 8 is the refinement factor, and $l$ is the AMR level of the grid patch.  We limit the refinement of the grid to ten levels throughout the run.

\subsection{Star Formation}
\label{SF}

We turn on both Pop III and Pop II SF in the halo at the beginning of each run.  The criteria for Pop III and Pop II SF are described in Sections 8.2.5 and 8.2.2 of \citet{enzo}, respectively.  The rollover from Pop III to Pop II SF occurs at a metallicity $Z =$ 10$^{-4}$ \Zs\ \citep[see, e.g.,][]{mbh03,schn06}.  If the conditions for Pop III SF are satisfied in a grid cell we assume for simplicity that just one star forms there.  In practice, we find that at most one or two Pop III stars form in a halo before it becomes enriched by metals.  Sampling the HIR15 IMF or some other logarithmically-flat distribution for the masses of these stars, as in \citet{wise12}, can therefore lead to highly disparate Pop III star masses from one halo to the next because of small-number statistics, making determination of the effect of Pop III IMF on the properties of the galaxies difficult.  We therefore assign any Pop III star that forms in a halo after the launch of the run only one of three masses: 20, 200 or 300 \Ms, which result in a CC SN, PI SN or collapse to a black hole (BH), respectively.  The Pop III star mass in a given halo thus becomes a parameter in our models.

We assume a SF efficiency $\epsilon_\mathrm{SF} =$ 0.05, in which 5\% of the mass of the gas in a cell that satisfies the criteria for SF is converted into a star.  At our grid resolution, at most $\sim$ 100 \Ms\ satisfies these criteria in a cell so the typical masses of the Pop II stars in our run are less than 10 \Ms.  The Pop II and III star particles in our runs therefore represent individual stars, not clusters, and we adopt a minimum mass of 1 \Ms\ for Pop II stars.

\subsection{SF Feedback}
\label{StellarFeedback}

We transport ionizing UV photons due to massive stars with the \moray\ ray-tracing package \citep{moray}. \moray\ includes radiation pressure on gas due to photoionizations and is self-consistently coupled to hydrodynamics and primordial gas chemistry in \enzo. Each star is taken to be a point source that emits both ionizing and Lyman-Werner (LW) photons.  We use mass-dependent luminosities and lifetimes for Pop III stars from \citet{s02} but calculate them for Pop II stars with 1D stellar evolution models with \texttt{MESA} \citep{paxt11,paxt13,paxt15,paxt18}. Ionizing photon rates and lifetimes for Pop II stars are tabulated at $10^{-2}, 10^{-3}$, and $10^{-4}$ \Zs\ and from 1 - 100 \Ms.  We neglect winds from Pop III stars because they are not thought to lose much mass over their lifetimes.  Mass-loss rates for Pop II stars vary as $\dot{M} \propto Z^{m}$ where $m \sim$ 0.1 - 1 for metallicities down to 0.01 \Zs\ \citep{vink01,nsmith14}. Since the metallicities of the Pop II stars in our models generally fall below this limit, where loss-rates are uncertain, and mass loss at this limit is modest, we also exclude winds from Pop II stars in our runs.  Because the focus of our work is to understand how metals from SNe affect the properties of primordial galaxies, we also neglect X-ray feedback due to BHs from the 300 \Ms\ Pop III stars in our halos.

\subsection{SN Feedback}

Energy from Pop II and Pop III SNe is deposited in the gas as thermal energy rather than linear momentum \citep[e.g.,][]{wet08a}.  This practice in principle can lead to the classic overcooling problem, in which large amounts of thermal energy deposited in high densities are radiated away by cooling before they can create the large pressure gradients that drive shocks outward, as in real explosions.  However, this is not a problem in our runs because, as discussed in the Introduction, ionizing UV flux from the star drives gas from it in strong supersonic flows and the explosions always occur in low densities \citep[stellar winds have a similar effect when present;][]{smidt18}.  Consequently, sound-crossing times in our models are always shorter than cooling times and SNe always drive strong shocks in our runs.

\section{Results}
\label{Results}

We performed 18 runs in which we varied the halo mass, the masses of the Pop III stars that form during the run, the number, type and spatial distribution of Pop III SNRs with which we initialize each halo, and the baryon density profile of the halo.  These models are partitioned into 6 groups, and what varies between models in a group is the mass of the Pop III stars that form in them.  The model parameters are summarized in Table~\ref{table:simic}.  SF rates in the galaxies in our models level off as ionizing UV and SNe due to stars begin to suppress the formation of new stars.  Since we are mostly interested in the transition from Pop III to Pop II SF (and computational costs rise as more stars form), we halt our simulations when SF flattens out and the properties of the stellar populations in the nascent galaxies stabilize. 

\begin{table}
\centering
\begin{tabular}{c c c c c c c c }
  Model & $M_\mathrm{halo}$ & SNR IMF & $\rho(r)$ & 
  $M_\mathrm{PopIII}$ & $N_\mathrm{CC}$ & $N_\mathrm{HN}$ & $N_\mathrm{PI}$\\
\hline
   A1    &   1.0e9   &   SAL       &   $r^{-1.0}$   &   20    &   18  &   2   &   0   \\
   A2    &   1.0e9   &   SAL       &   $r^{-1.0}$   &   200  &   18  &   2   &   0   \\
   A3    &   1.0e9   &   SAL       &   $r^{-1.0}$   &   300  &   18  &   2   &   0   \\
\hline
   B1    &   1.0e9   &   HIR15   &   $r^{-1.0}$   &   20    &   2    &   0   &   7   \\
   B2    &   1.0e9   &   HIR15   &   $r^{-1.0}$   &   200  &   2    &   0   &   7   \\
   B3    &   1.0e9   &   HIR15   &   $r^{-1.0}$   &   300  &   2    &   0   &   7   \\
\hline
   C1    &   1.0e9   &   SAL       &   $r^0$         &   20    &   18  &   2   &   0   \\
   C2    &   1.0e9   &   SAL       &   $r^0$         &   200  &   18  &   2   &   0   \\
   C3    &   1.0e9   &   SAL       &   $r^0$         &   300  &   18  &   2   &   0   \\    
\hline
   D1    &   1.0e9   &   HIR15   &   $r^0$          &   20    &   2    &   0   &   7   \\
   D2    &   1.0e9   &   HIR15   &   $r^0$          &   200  &   2    &   0   &   7   \\
   D3    &   1.0e9   &   HIR15   &   $r^0$          &   300  &   2    &   0   &   7   \\
\hline
   E1    &   3.0e8   &   SAL       &   $r^{-0.1}$   &   20    &   5    &   0   &   0   \\
   E2    &   3.0e8   &   SAL       &   $r^{-0.1}$   &   200  &   5    &   0   &   0   \\
   E3    &   3.0e8   &   SAL       &   $r^{-0.1}$   &   300  &   5    &   0   &   0   \\   
\hline
   F1    &   3.0e8   &   HIR15   &   $r^{-0.1}$   &   20     &   0    &   0   &   3  \\
   F2    &   3.0e8   &   HIR15   &   $r^{-0.1}$   &   200   &   0    &   0   &   3  \\
   F3    &   3.0e8   &   HIR15   &   $r^{-0.1}$   &   300   &   0    &   0   &   3  \\
  \hline
\end{tabular}
\caption{Model parameters. Left to right: model, halo mass (\Ms), Pop III SNR IMF, baryon density profile, Pop III star mass (\Ms), number of CC SNRs, number of HN SNRs, and number of PI SNRs.}
\label{table:simic}
\end{table}

\begin{figure*} 
\centering
\includegraphics[width=0.9\textwidth]{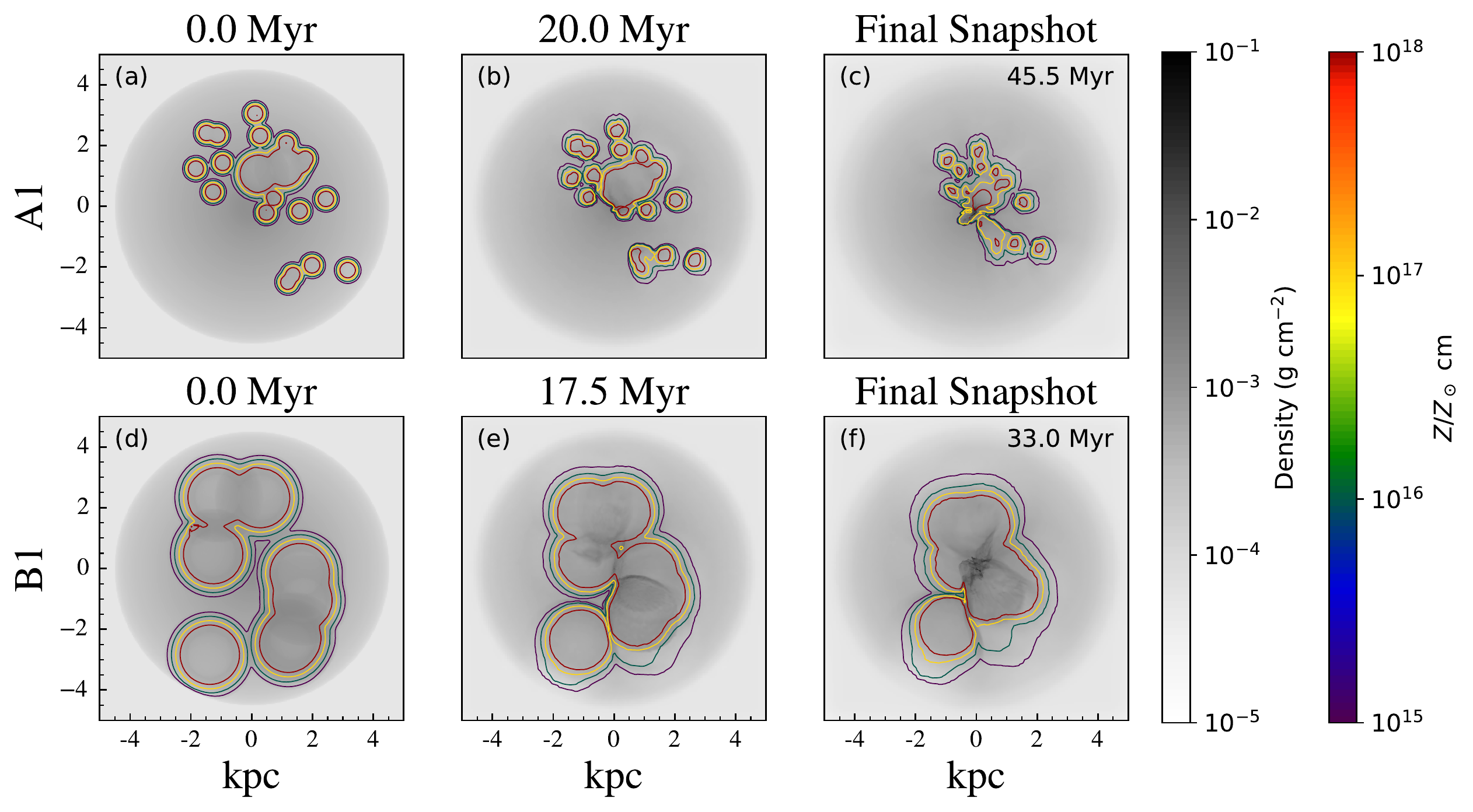}
\caption{Density projections at the beginning, intermediate, and final stages of the run, with corresponding times shown in each panel. Density-weighted metallicities appear as colored contours. The top and bottom panels are for A1 and B1, respectively.  Mixing of Pop III SNRs with primordial gas can be seen as the increasing separation between some contours over time.}
\label{fig:rgasmetala}
\end{figure*}

\begin{figure*}
\centering
\includegraphics[width=0.9\textwidth]{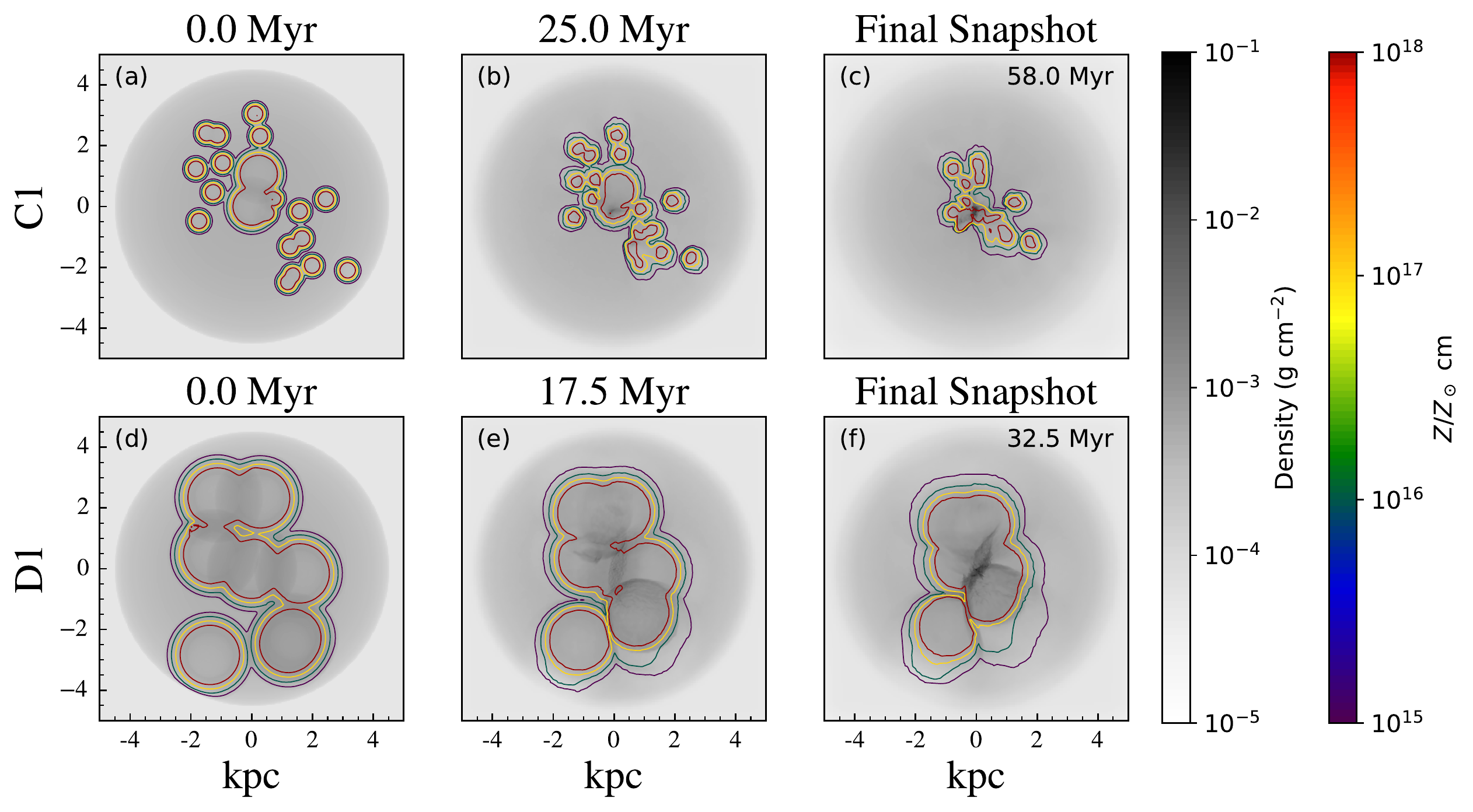}
\caption{Same as Fig.~\ref{fig:rgasmetala} but for C1 and D1.}
\label{fig:rgasmetalb}
\end{figure*}

\subsection{Gas Flows}
\label{rgasdynamics} 

\begin{figure*}
\centering
\includegraphics[width=0.9\textwidth]{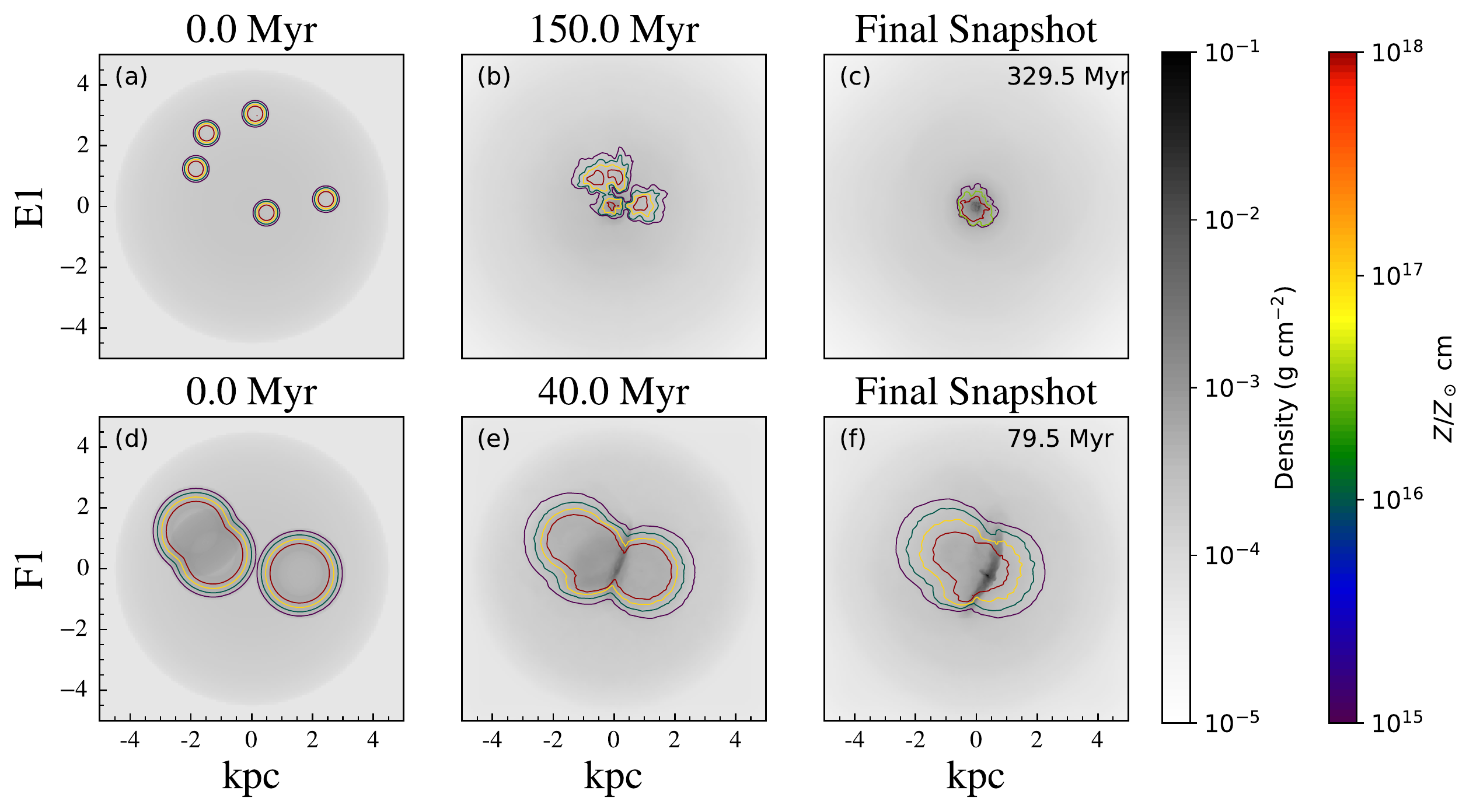}
\caption{Same as Figs.~\ref{fig:rgasmetala} and \label{fig:rgasmetalb} but for E1 and F1.  All the SNRs have merged at the center of the halo in E1 because of its longer evolution times. }
\label{fig:rgasmetalc}
\end{figure*}

In Figures~\ref{fig:rgasmetala}--\ref{fig:rgasmetalc} we show projections of density and metallicity in A1, B1, C1, D1, E1, and F1 (only the first model in each group is shown because the evolution of galaxies within a group is similar).  C1 and D1 can be distinguished from the others by their flat density profiles, which lack a gradient in grayscale towards their centers.  From the metallicity contours it is evident that the SNRs migrate to the center of the halo over time and that they retain their basic morphologies until they begin to collide.  The mixing of metals in SN bubbles with ambient pristine gas is visible as the increasing separation between some of the metallicity contours over time.  The distinct halo boundaries at 4 kpc become blurred over time as gas flows erase the sharp drop in density there.  Atomic cooling in the halos results in a nearly isothermal equation of state for the gas in which sharp density gradients become strong pressure gradients.  These pressure gradients cause gas at the edge of the halo to expand outward, smoothening out truncated densities there.

Evolution times are much longer in the less massive halo because of its weaker gravitational field, so SNRs fall to the center more slowly.  The final distribution of remnants in the halos varies significantly across the runs, from occupying much of the volume of the halo in D1 to being confined to small radii at the center in E1.  The range of filling factors is due to the energy of the explosions:  in the B1, D1 and F1 models the much larger PI SN remnants are easily distinguished from the less energetic CC SNRs in the other runs.  Metals produced by new Pop III stars mostly only appear in the central 200 pc.  

\begin{figure*}
\centering
\includegraphics[width=0.8\textwidth]{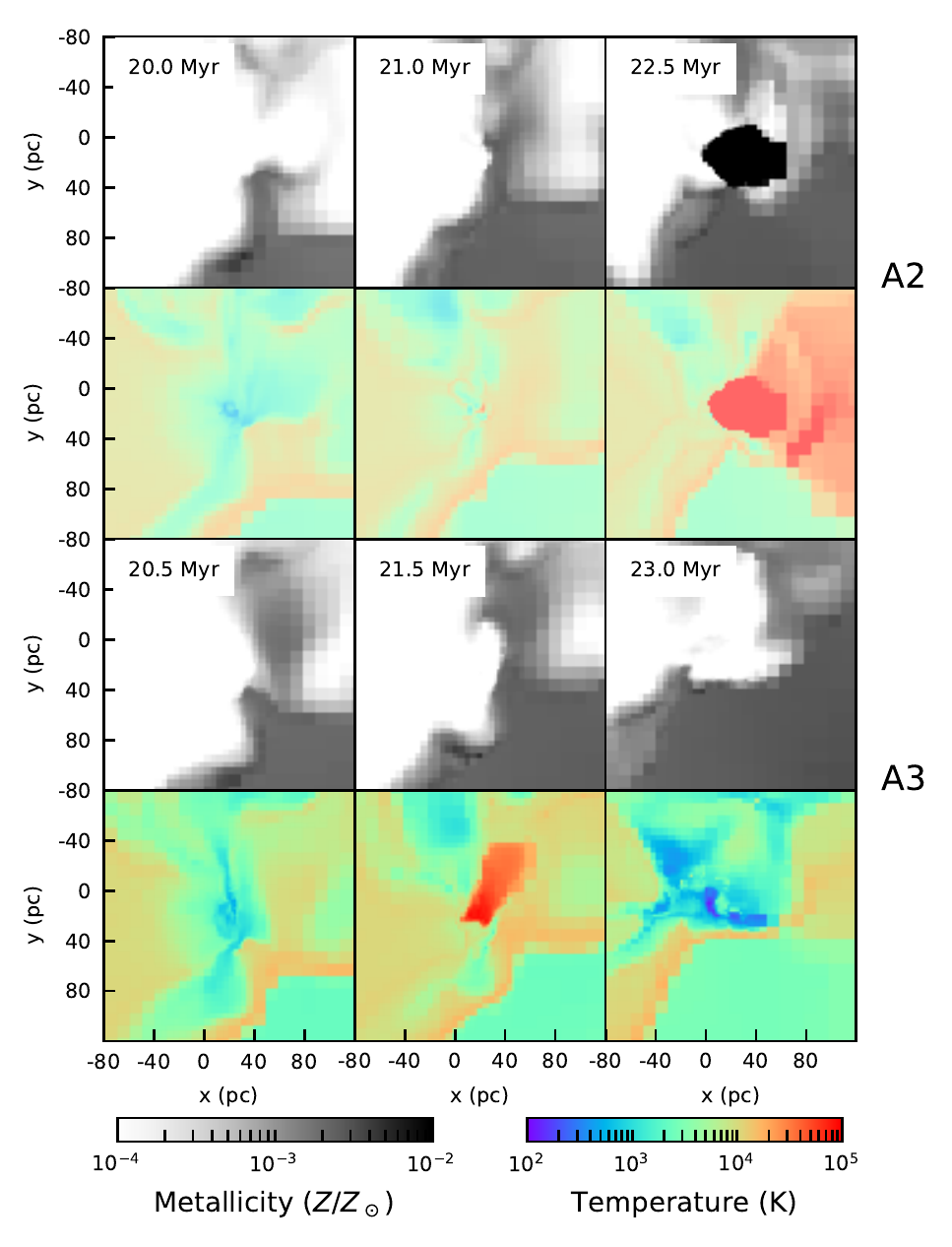}
\caption{Metallicity and temperature slices of the central 200 pc, where a new Pop III star forms in A2 and A3.  Left to right:  0.5 Myr before birth, 1 Myr after birth, and 0.5 Myr after death.  Radiative feedback ie evident in the temperatures at 1 Myr.  The metals in the left and middle panels are due to the original Pop III SNRs but new metals blown out by the PI SN are visible in the top right panel.}
\label{fig:rmetaltemperature}
\end{figure*}

We show radiative feedback and metal injection due to these stars in A2 and A3 in Fig.~\ref{fig:rmetaltemperature}.  In A2 the star dies as a PI SN and in A3 it collapses to a BH.  In both cases the Pop III star heats and ionizes surrounding gas, which can either suppress or promote new SF in its vicinity \citep{wet08b,wet10}.  In A2 the star dies as a PI SN, heating the gas to high temperatures ($> 10^{5}$ K) and ejecting a large mass of metals that enhance cooling and promote a transition to Pop II SF.  Because we neglect X-rays from the BH the gas cools down quickly after the death of the star in A3.  The gas recombines out of equilibrium, as  it remains partially ionized even after its temperature has fallen to only a few thousand K.  This free electron fraction catalyzes H$_2$ formation via the H$^-$ channel that cools gas in the relic H II region of the star down to approximately 150 K, as seen in the image on the right.  In reality, X-rays would ionize and heat the gas out to larger radii than that of the H II region of the star and, depending on its spectrum, enhance free electron fractions there via secondary ionizations by energetic photoelectrons.  These fractions would also promote H$_2$ formation, cool the gas, and trigger more star formation \citep[e.g.,][]{mba03,aycin14,smidt18}.  We therefore likely somewhat underestimate the true rates of SF in our simulations. 

\begin{figure*}
\centering
\includegraphics[width=0.7\textwidth]{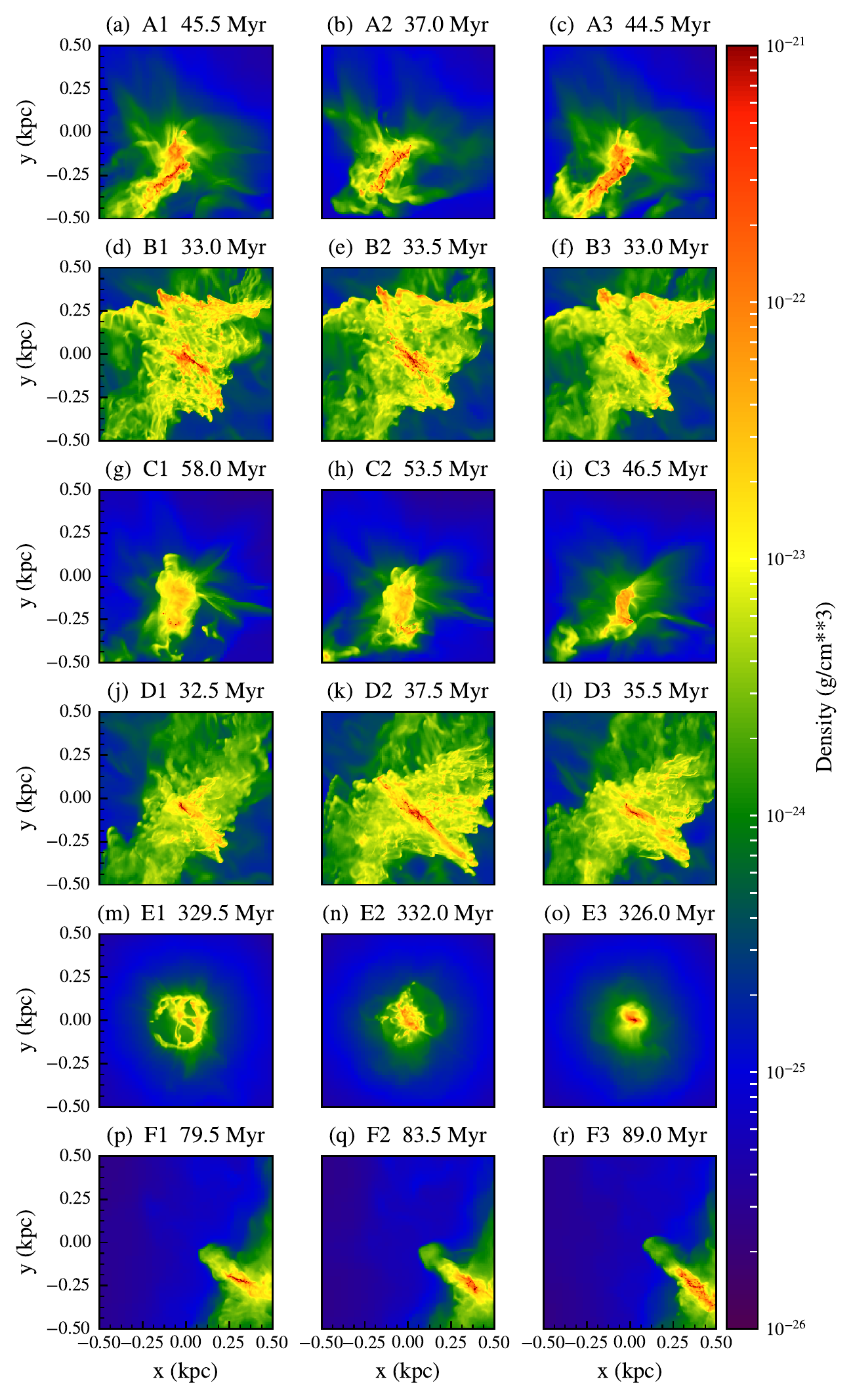} 
\caption{Density slices of the central 1 kpc at the end of each run.}
\label{fig:rgaszooma}
\end{figure*}

We show slices of final gas density at the center of each halo in Figure~\ref{fig:rgaszooma}.  Collisions between SNRs produce a range turbulent flows at the center, from fairly violent ones in the B and D series to more quiescent ones in the E series.  Densities in the A and C series are less turbulent because they are the product of collisions between CC SNRs, not the PI SNRs in B and D.  Final densities in the E series are nearly spherical because they had the fewest SNRs and the remnants have had much more time to settle at the center of the halo than in the other runs.  The regions of highest density in the F series are displaced $\sim$ 200 pc from the center of the halo because of asymmetries in the initial distribution of the SNRs.  

\begin{figure*}
\centering
\includegraphics[width=0.8\textwidth]{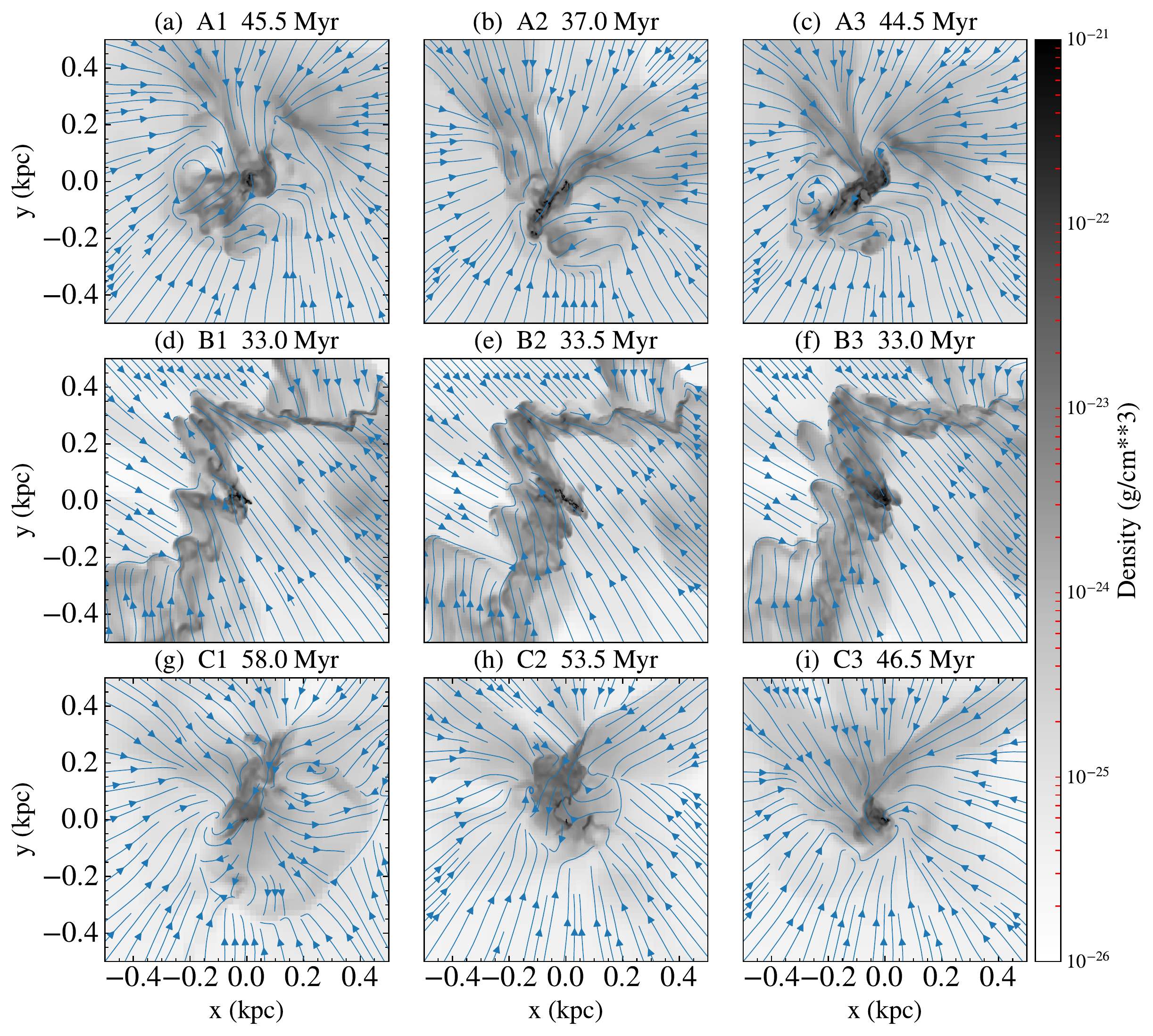}
\caption{Density slices of the central 1 kpc at the end of the A1 -- C3 runs.  The blue streamlines mark the directions (but not velocities) of gas flows.}
\label{fig:rgasstreeama}
\end{figure*}

\begin{figure*}
\centering
\includegraphics[width=0.8\textwidth]{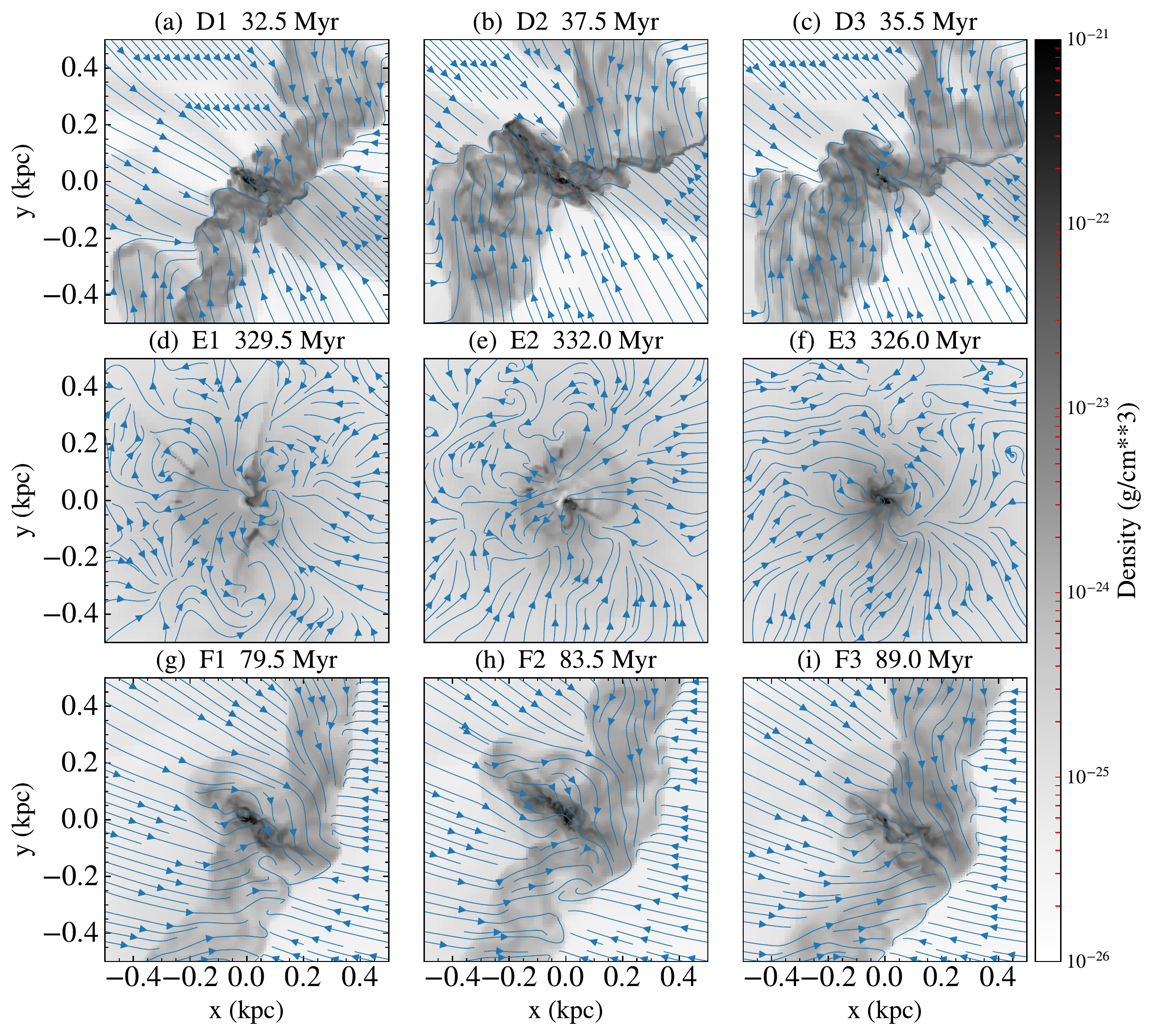}
\caption{Same as Fig.~\ref{fig:rgasstreeama} but for the D1 -- F3 runs.}
\label{fig:rgasstreeamb}
\end{figure*}

These structures are not strongly affected by SNe of new Pop III stars because there are no such explosions in the D series but they have morphologies similar to those in the B series, which do have new Pop III SNe.  Figures~\ref{fig:rgasstreeama} and \ref{fig:rgasstreeamb} show streamlines at the center of the halo at the end of each run.  In the SAL models (A1, C1, and E1), the gas is flowing into the center from all directions, although in E1 there are eddy-like structures that indicate turbulence.  In the HIR15 models (B1, D1, and F1), there are discontinuities in flows due to collisions between SNRs during the merging process. 

\subsection{Metal Fractions}

\begin{figure*}
\centering
\includegraphics[width=0.9\textwidth]{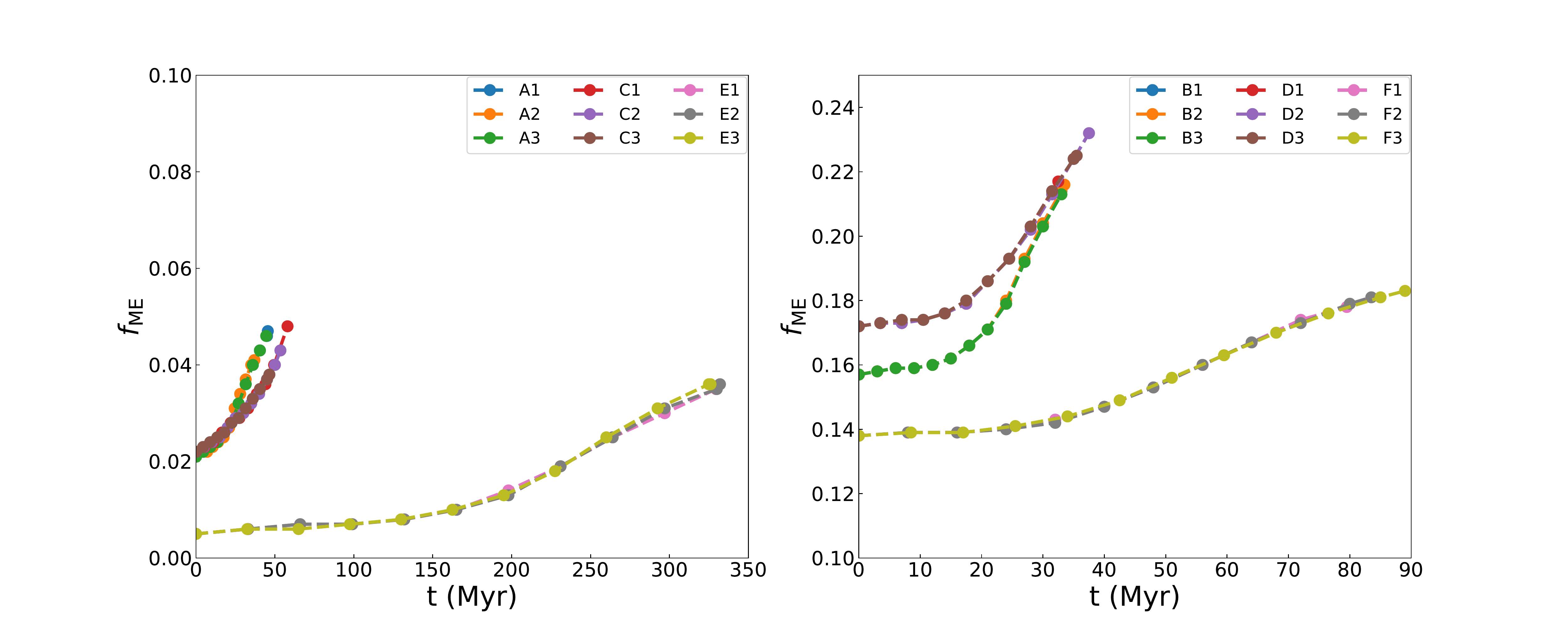}
\caption{Evolution of the fraction of metal-enriched gas, $f_{\mathrm{ME}} = M_\mathrm{ME} / M_\mathrm{gas}$, in the 18 runs. $M_\mathrm{ME}$ is the total gas mass at $Z_\mathrm{gas} > 10^{-4}~Z_\odot$. Metallicities rise at earlier times in the more massive halo because the Pop III SNRs fall to its center and mix with pristine gas along the way more quickly.  The A series metallicities grow more rapidly than those in the C series, and the B series metallicities rise more quickly than in the D series because new Pop III stars inject extra metals into the halos when they die as SNe.}
\label{fig:rmetalenricha}
\end{figure*}

Mass fractions for metal-enriched gas versus time for all 18 models are shown in Figure~\ref{fig:rmetalenricha}.  Here, $f_\mathrm{ME} = M_\mathrm{ME} / M_\mathrm{gas}$, where $M_\mathrm{ME}$ is the total gas mass with $Z_\mathrm{gas} > 10^{-4}~Z_\odot$ (the same criterion for Pop II SF).  In the left panel are models with the SAL IMF for Pop III SNRs and in the right panel are those with the HIR15 IMF.  First, we emphasize that the rise in metallicity out to 30 - 40 Myr in the A - D series is purely due to mixing of Pop III SNRs with primordial gas in the halo (and metals from the few new Pop III stars), not to Pop II SF, which begins after 30 Myr in all 18 models as we discuss in the next section.  The rise in metallicity is gradual at first in all the runs but then steepens as the SNRs accelerate towards the center of the halo and more turbulently mix with ambient gas.  The B, D and F models start out with a much higher $f_\mathrm{ME}$ than the A, C and E models because the PI SNRs contain far more metals than the CC SNRs.  

The D and C models have higher initial metallicity fractions than the B and A models because a Pop III SNR was initialized at the center of the halo and encloses some of its flat density profile.  The difference between A and C is much smaller than between B and D because the bubble at the center of the halo is due to a CC SN, not a PI SN.  The A series metallicities grow more rapidly than in the C series, and the B series metallicities rise more quickly than in the D series, because new Pop III PI SNe inject more metals into the halos than new Pop III CC SNe.  The E and F series have the lowest initial metallicity fractions because they begin with the fewest SNRs, and they are the slowest to rise because the SNRs require more time to reach the center of the halo because of its shallower potential well.  The evolution in metallicity within members of a given series is virtually identical because they differ only in the masses assumed for new Pop III stars.  Since at most two such stars form in each model, as discussed below, their contribution to the enrichment of the halo is small in comparison to that of the original Pop III SNRs.  

\begin{figure*}
\centering
\includegraphics[width=0.8\textwidth]{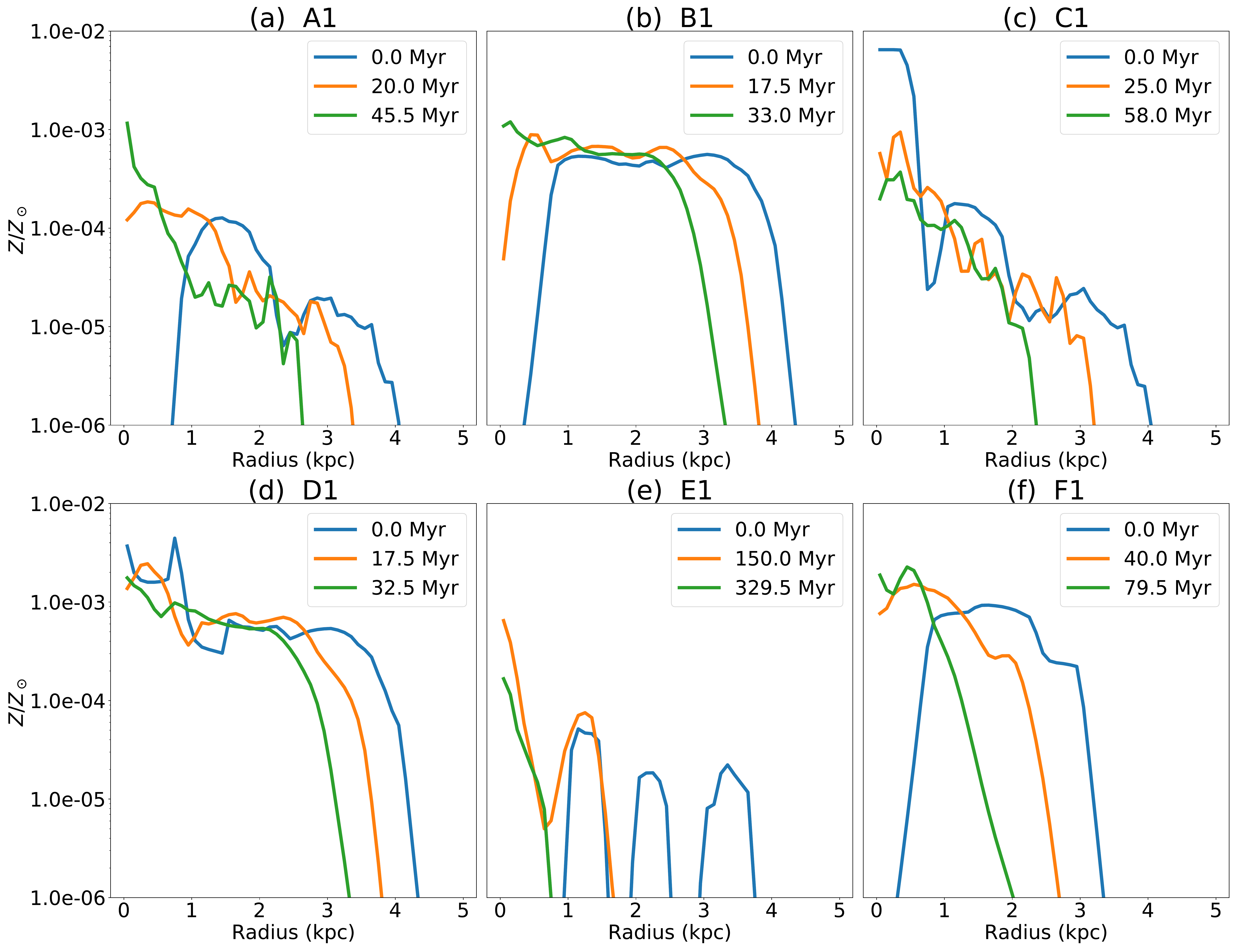}
\caption{Spherically-averaged metallicities at early, intermediate and late times in A1 - F1.}
\label{fig:rradialmetala}
\end{figure*}

\begin{figure*}
\centering
\includegraphics[width=0.8\textwidth]{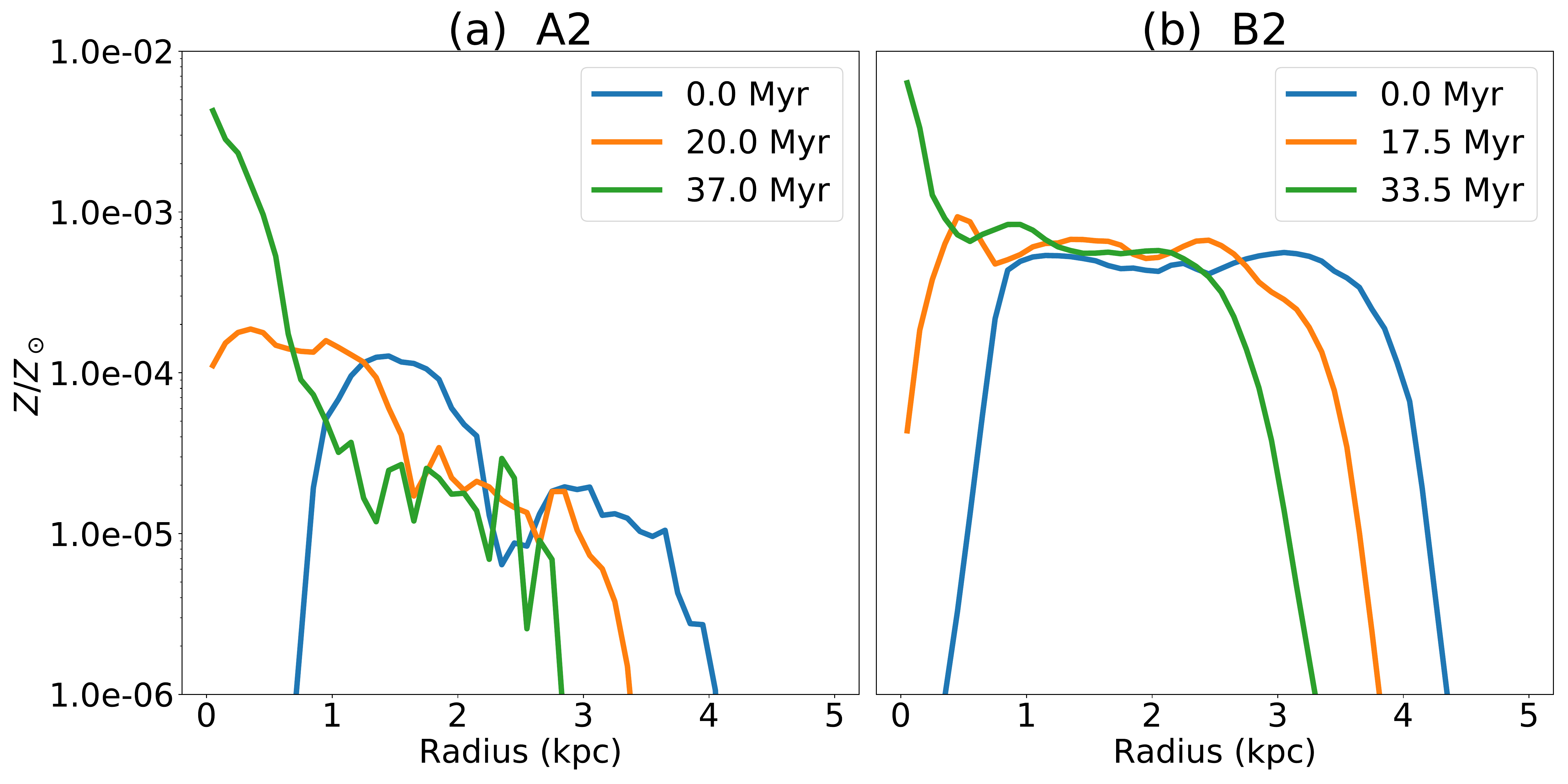}
\caption{Spherically-averaged metallicities at early, intermediate and late times in A2 and B2.}
\label{fig:rradialmetalb}
\end{figure*}

We show spherically-averaged metallicities for the A1 - F1 runs in Figure~\ref{fig:rradialmetala}.  In A1, B1, E1 and F1, the pooling of metals in SN bubbles at the center of the halo is evident as metallicities there rise over time.  In contrast, metallicities in the center of the halo are greatest at early times in the C1 and D1 runs because a SNR is initialized there.  Central metallicities are higher in C1 than in D1 because the PI SNR in D has driven metals out to larger radii.  Metallicities at the center of the halo in these two runs fall over time because high temperatures in the SN bubble cause it to expand out into the halo and because pristine gas is also falling towards the center of the halo.  The ripples in metallicity at intermediate and late times in A1 - D1 are due to inhomogeneities in mixing that arise during collisions between SNRs.  These fluctuations have smaller amplitudes in B1 and D1 because there is more turbulence in the halo that smoothens out inhomogeneities, as seen in Figure~\ref{fig:rgaszooma}.  We show radial metallicity profiles for A2 and B2 in Figure~\ref{fig:rradialmetalb}.  These two runs form new Pop III stars that die as PI SNe instead of the CC SNe in A1 and B1.  A2 and A1 have similar metallicities at early and intermediate times but A2 has much higher central metallicities at late times because of the explosion of the 200 $M_\odot$ star.  B1 and B2 follow similar trends. 

\subsection{Star Formation}
\label{rngos} 

\begin{table}
\centering
\begin{tabular}{c c c c c c c c}
Model & $t_\mathrm{evol}$ (Myr) & $N_\mathrm{PopII}$ & $M_\mathrm{*}$ ($M_\odot$) & $N_\mathrm{PopIII}$ & $\alpha$ \\ 
\hline
   A1   &   45.5     &   1742     &   3350     &   2   &   1.81   \\
   A2   &   37.0     &   2022     &   3966     &   1   &   1.82   \\
   A3   &   44.5     &   2911     &   4412     &   1   &   3.19   \\ 
\hline
   B1   &   33.0     &   3031     &   3816     &   1   &   4.76   \\
   B2   &   33.5     &   1968     &   2655     &   1   &   3.58   \\
   B3   &   33.0     &   683       &   1174     &   2   &   1.66   \\
\hline
   C1   &   58.0     &   170       &   225       &   0   &   2.76   \\
   C2   &   53.5     &   234       &   391       &   0   &   1.78   \\
   C3   &   46.5     &   756       &   1576     &   0   &   1.46   \\   
\hline
   D1   &   32.5     &   2295     &   3409     &   0   &   2.78   \\
   D2   &   37.5     &   1241     &   1692     &   0   &   3.38   \\
   D3   &   35.5     &   6741     &   13803   &   0   &   1.68   \\
\hline
   E1   &   329.5   &   822       &   1141     &   0   &   3.57   \\
   E2   &   332.0   &   1324     &   2667     &   0   &   1.37   \\
   E3   &   326.0   &   3552     &   6863     &   0   &   1.64   \\   
\hline
   F1   &   79.5     &   9085     &   18761   &   0   &   1.91   \\
   F2   &   83.5     &   5500     &   9397     &   0   &   2.22   \\
   F3   &   89.0     &   4525     &   9972     &   0   &   1.57   \\
\hline
\end{tabular}
\caption{Final stellar populations in all 18 models.  Left to right:  model, halo evolution time, final number of Pop II stars, final total stellar mass, number of new Pop III stars forming at any time in the run, and $\alpha$ (where d$N$/d$M_\mathrm{star}$ $\propto M_\mathrm{star}^{-\alpha}$).}
\label{table:rmainresults}
\end{table}

\begin{figure*}
\centering
\includegraphics[width=0.9\textwidth]{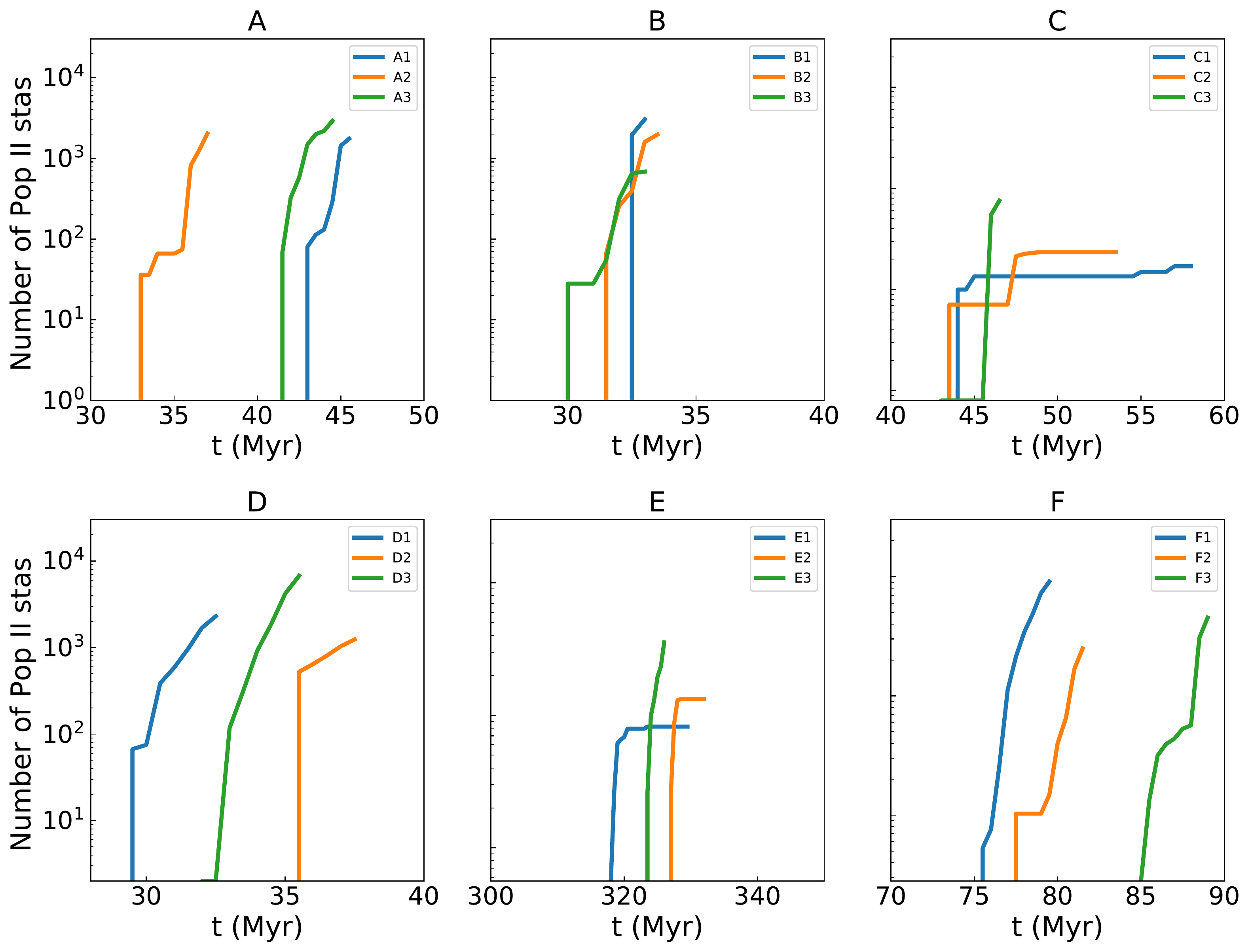}
\caption{Pop II star formation in all 18 runs. SF rates level off because ionizing UV flux and SN blast waves begin to suppress the formation of new stars.}
\label{fig:msfrh}
\end{figure*}

We plot Pop II SF rates in Figure~\ref{fig:msfrh} and tally final numbers of Pop III and Pop II stars for all 18 models in Table \ref{table:rmainresults}.  All Pop II stars form within the central 200 pc of the halo except in the F series, in which they are confined to a 200 pc region that is offset from the center by about 200 pc, as discussed in Section~3.1.  New Pop III stars only form in the A and B series because central densities in the deeper potential well of the more massive halo are high enough to trigger SF before the gas is polluted by SNRs.  In groups C and D, gas in the flat profiles cannot collapse to high enough densities to form stars before it is polluted by SNRs (and gas at the center of the halo has already been polluted by one remnant).  Pop II SF begins at similar times in the A - D runs but much later in the E and F models because free-fall times for the SNRs to the center of the less massive halo are much longer, so more time is required for metals to enrich gas that is at high enough densities to form new stars.  These two series do not produce new Pop III stars because pristine gas never reaches high enough densities in the shallower potential well of this halo to form stars before being chemically enriched, even in the $r^{-1}$ profiles.  The total stellar mass at the end of the runs is $\sim 10^3$ -- $10^4$ $M_\odot$. 

\begin{figure*}
\centering
\includegraphics[width=0.7\textwidth]{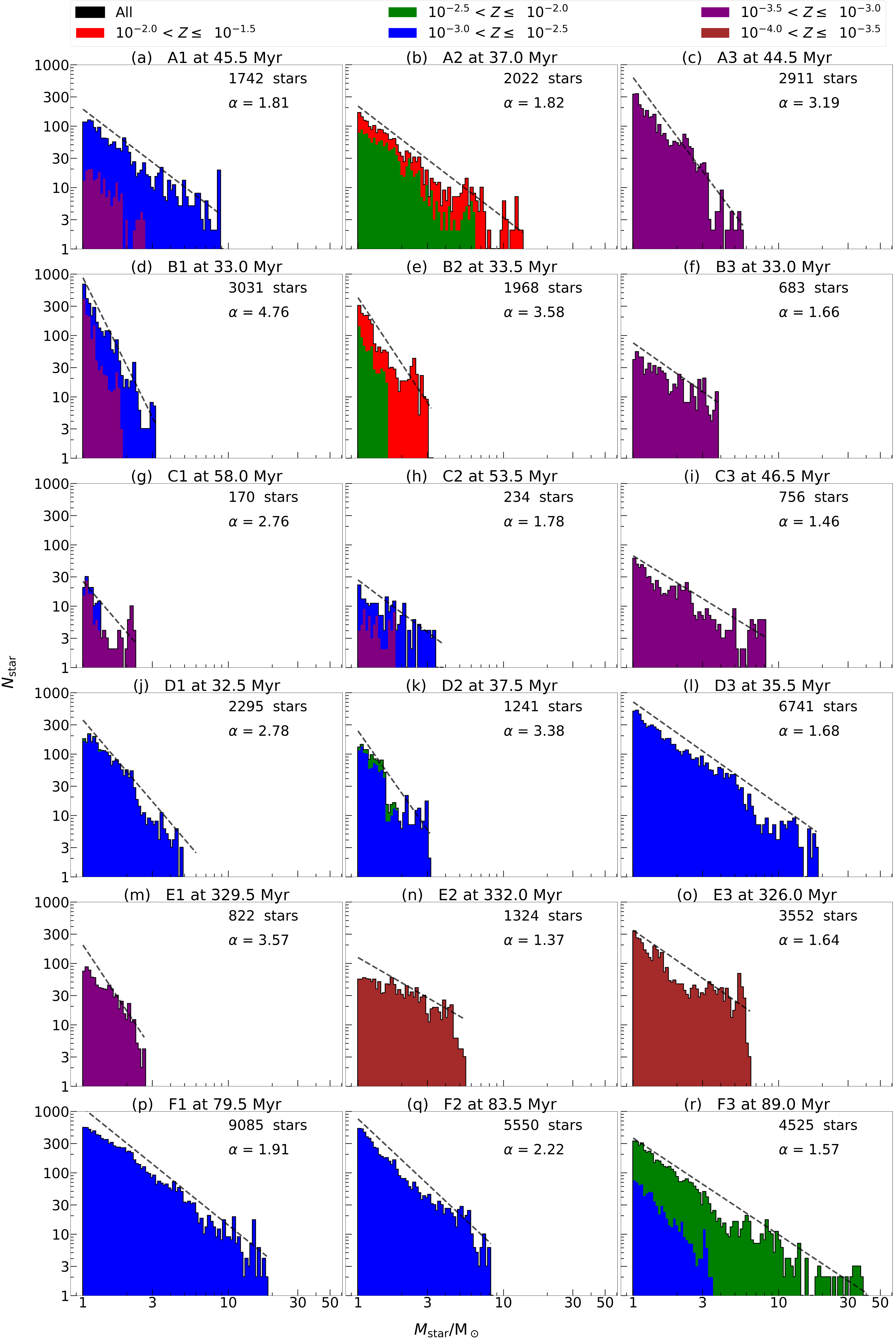}
\caption{Final Pop II mass distributions in all 18 models. The total number of stars is shown in the upper right corner of each panel.  No Pop III stars remain at the end of any of the runs. The stars are partitioned according to metallicity (10$^{-4}$ -- 10$^{-1.5}$ $Z_\odot$).  They fall into a power-law distribution in mass in most of the runs except for C1 and C2, in which there are too few stars to construct a simple distribution.  A second peak is present at the high end of the mass range in A2, C3, D1, and D3.  The black dashed line represents the best fit power-law function for each population.  We only use the first peak in D1 and D3 to compute the power-law fit.  The best fit exponent $\alpha$ is shown in the upper right corner of each panel.}
\label{fig:rsm}
\end{figure*}

We show Pop II mass functions in Figure~\ref{fig:rsm}, color-coded by metallicity.  These are the masses and metallicities the SF algorithm assigns to each star particle based on the gas mass in a given cell(s) that satisfies the criteria for SF.  Most of the distributions can be fit by a power-law distribution, d$N$/d$M_\mathrm{star}$ $\propto M_\mathrm{star}^{-\alpha}$. The most massive stars in all the models are $\sim 35$ $M_\odot$.  In D1 and D3 there is a second peak at the high mass end.  We find that $\alpha$ varies from 1.37 -- 4.76 across all 18 models and is correlated with the degree of turbulence at the sites of SF, with high-mass SF falling off sharply in more turbulent gas.  Metallicities in the A and B models are highest in the 2-series because new Pop III stars die as PI SNe and most heavily enrich the gas from which Pop II stars form.  The next highest metallicities are in the 1-series, in which Pop III CC SNe explode, followed by the 3-series, in which new Pop III stars simply collapse to BHs and the only metals are due to the SNRs inserted at the beginning of the run.  Pop II star metallicities in the C - F runs are lower because no new Pop III stars form, so metals come only from SNRs.  They are higher in the D and F runs because the SNRs are high-yield PI SNe.

Pop II SF begins later in SAL halos than in HIR15 halos because fewer metals result in less efficient cooling so gas collapses into stars at later times.  In 10$^9$ \Ms\ halos with uniform initial densities (Figures~\ref{fig:rsm}g -- l), the mass functions are more irregular, with a second peak appearing at the high mass end.  Metallicities in the second peak are similar to those in the first.  In these cases, low-mass stars formed first, creating the first peak.  Then radiative feedback from these stars heated nearby clouds and increased their Jeans masses.  When gas in these clouds collapses into more massive stars the second peak forms.   

\begin{figure*}
\centering
\includegraphics[width=0.7\textwidth]{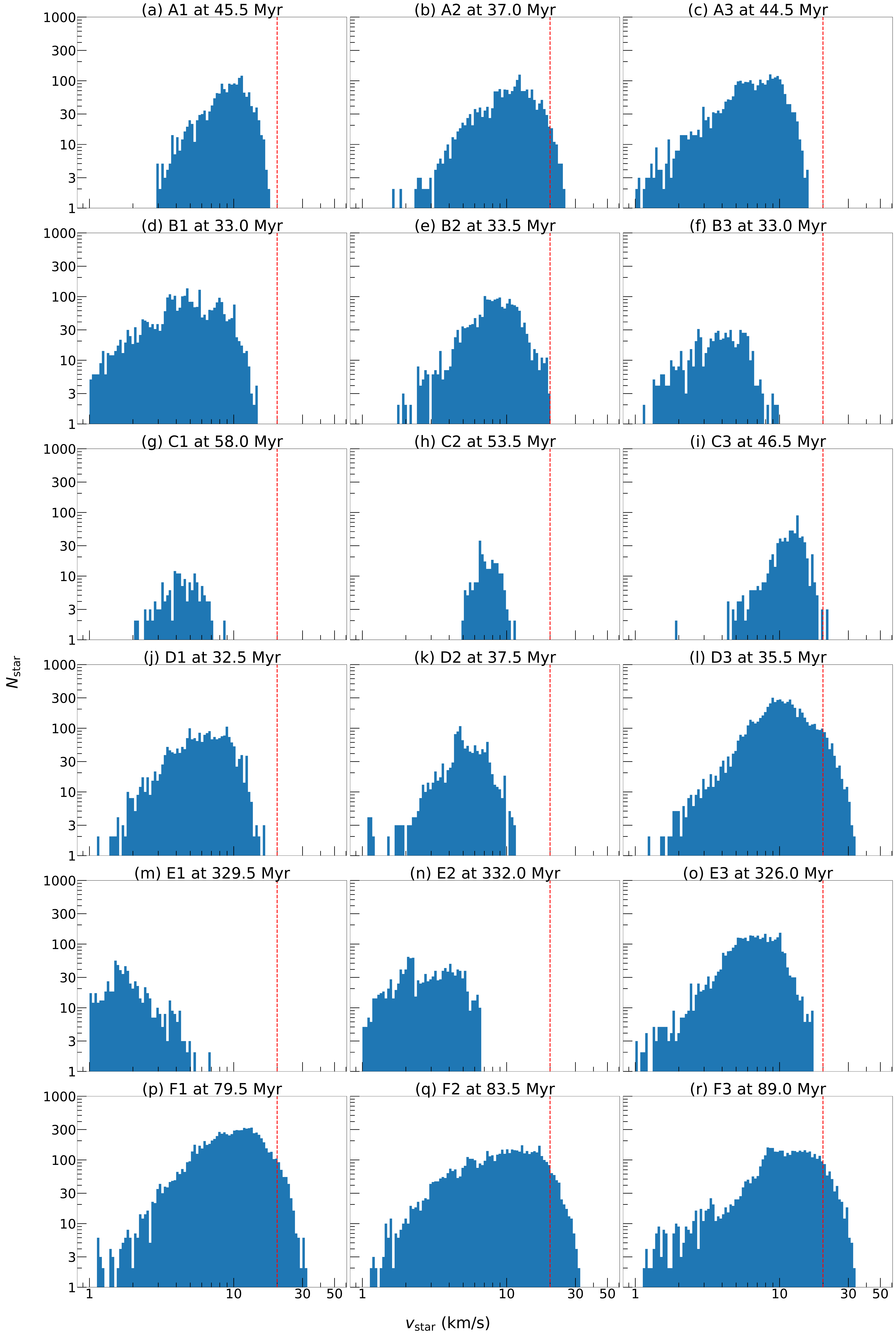}
\caption{Stellar velocity distributions at the end of each run.  In D1, F1 and F2 more than 10\% of the stars have velocities above 20 km s$^{-1}$ (vertical red line) and could later escape the halos.} 
\label{fig:rsv}
\end{figure*}

\begin{figure*}
\centering
\includegraphics[width=0.8\textwidth]{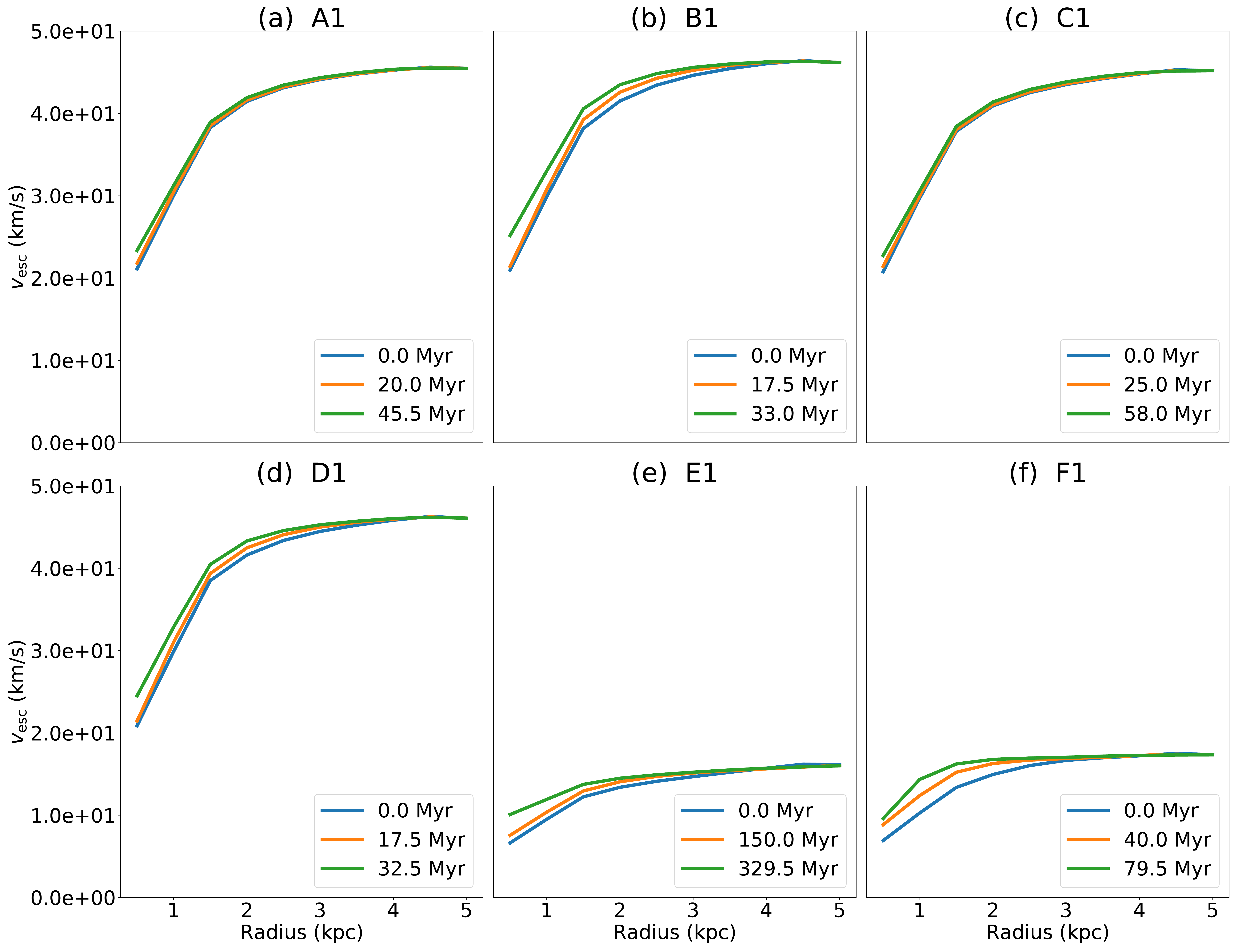}
\caption{Escape velocity as a function of the mass enclosed by the given radius at three stages of galaxy evolution. Only the first model of each group is shown because members of a given group have similar velocity profiles.} 
\label{fig:rkv}
\end{figure*}

Final Pop II star velocity distributions are shown in Figure~\ref{fig:rsv}.  Most of the stars have velocities of a few km s$^{-1}$ but a few have velocities greater than 30 km s$^{-1}$.  We show escape velocity from the halo versus radius for six models at several stages of evolution in Figure~\ref{fig:rkv}.  The escape velocity is v$_\mathrm{esc} = \sqrt {\mathrm{2G}M/R}$, where $M$ is the mass enclosed at a given radius, and is 40 - 50 km s$^{-1}$ in A - D and 15 - 20 km s$^{-1}$ in E and F.  Some of the stars in E and F could therefore escape the galaxy if it is allowed to evolve longer.   

\begin{figure*}
\centering
\includegraphics[width=\textwidth]{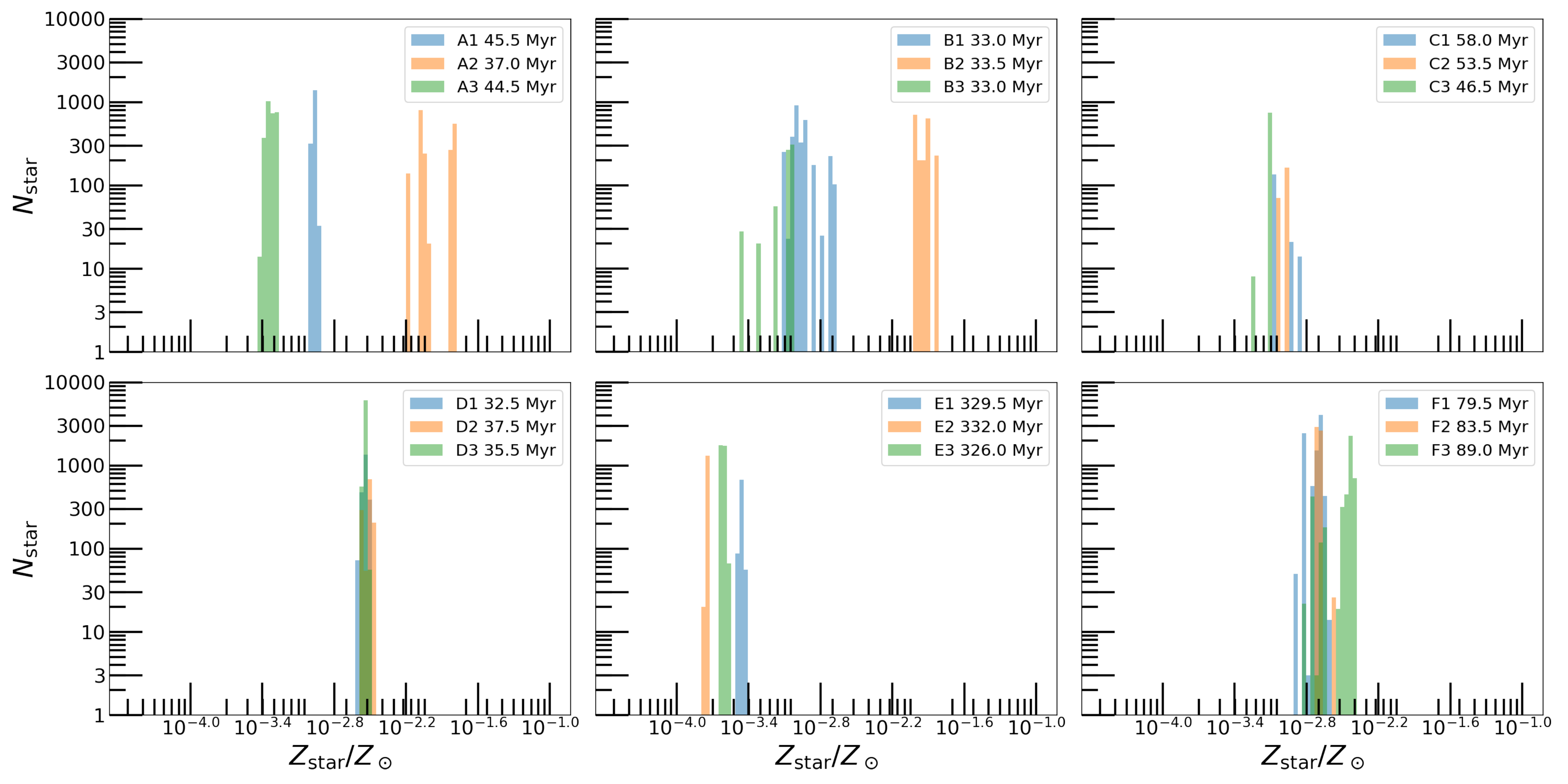}
\caption{Pop II stars in each galaxy binned by metallicity at the end of the run.}
\label{fig:dstellarmetal}
\end{figure*}

We bin Pop II stars by metallicity in Figure~\ref{fig:dstellarmetal}.  Stars in the A2 and B2 runs have the highest metallicities because new Pop III stars die as PI SNe and seed them with the most metals.  A1 and B1 have the next highest metallicities (being enriched by Pop III CC SNe, not PI SNe) and A3 and B3 have the lowest metallicities because new Pop III stars collapse without producing any metals.  Overall, the B series has higher metallicities than the A series because its SNRs are PI SNe.  The C series metallicities are much tighter because no new Pop III stars form and enrich the gas, so the variations in metallicity in the three runs are due to variations in the initial distributions of SNRs in the halos.  The same is true of the D series except the stars have higher metallicities because the SNRs are PI SNe.  However, there is even less variation in metallicity in the D runs because the gas at the center of the halo is more turbulently mixed by collisions between PI SNRs there (as seen in the D and C panels in Figure~\ref{fig:rgaszooma}).  The E series produce the fewest metals because they are initialized with just five CC SNRs.  Stellar populations in these three runs have distinct metallicities because collisions between SNRs at the center of the halo do not produce much turbulence.  There is much more overlap in metallicity in the F series because collisions between remnants results in more turbulent mixing, and the final metallicities are higher than in the E series because the F runs were initialized with PI SNRs.

\section{Conclusion}
\label{Conclusion}

We find that the Pop III IMF has a significant impact on the stellar populations of primordial galaxies, and thus their observational signatures and prospects for detection by {\em JWST} and extremely large telescopes (ELTs) on the ground in the coming decade.  Populations arising in the debris of very massive Pop III SNRs in general have higher metallicities but steeper mass functions than those forming primarily in less massive, Salpeter-like Pop III SNRs.  This trend is primarily due to the fact that collisions between PI SNRs in the small galaxy at birth drive more turbulence that tends to suppress high-mass star formation and because PI SNe produce nearly 100 times the metals of CC SNe and more than 10 times the metals of HNe.  

Our models also suggest that new Pop III stars were not a major component of the earliest galaxies because gas in all but the most massive halos was usually polluted by metals from SNRs during hierarchical assembly before it could collapse into pristine stars.  We find that only more massive galaxies form Pop III stars regardless of IMF because primordial gas can pool to higher densities in their deeper potential wells and collapse into stars before metals reach them.  Stellar populations arose at earlier times in primordial galaxies if the Pop III IMF was top-heavy because gas in halos became contaminated by metals at earlier times, cooling and forming stars.  

In principle, the Pop III IMF could be inferred from detections of primordial galaxies in the coming decade.  Top-heavy IMFs result in less massive Pop II stars and dimmer galaxies at early times than do more conventional Salpeter masses for the first stars.  In future work we will post process these simulations to derive synthetic observables for primeval galaxies, which may soon be detected by the next generation of space missions and telescopes on the ground.

\acknowledgments

We thank Hiroyuki Hirashita and You-Hua Chu for useful comments. L.-H. C. acknowledges financial support from the German Research Foundation (DFG) via the Collaborative Research Center (SFB 881, Project-ID 138713538) 'The Milky Way System' (subprojects A1). K. C. was supported by an EACOA Fellowship and by the Ministry of Science and Technology, Taiwan, R.O.C. under Grant no. MOST 107-2112-M-001-044-MY3.  He also thanks the Aspen Center for Physics, which is supported by NSF PHY-1066293, and the Kavli Institute for Theoretical Physics, which is supported by NSF PHY-1748958.  D. J. W. was supported by the Ida Pfeiffer Professorship at the University of Vienna.  Numerical simulations were performed at the National Energy Research Scientific Computing Center (NERSC), a U.S. Department of Energy Office of Science User Facility operated under Contract No. DE-AC02-05CH11231, at the Center for Computational Astrophysics (CfCA) at the National Astronomical Observatory of Japan (NAOJ), and at the TIARA Cluster at the Academia Sinica Institute of Astronomy and Astrophysics (ASIAA).  The analysis and plots were done with \texttt{yt} \citep{yt}.


\bibliographystyle{yahapj}
\bibliography{refs}


\end{document}